\documentclass[namedreferences]{solarphysics}
\usepackage{multirow}
\usepackage{rotating}
\usepackage{longtable}
\usepackage{tabularx}
\usepackage[hyperref,optionalrh,solaromanenum]{spr-sola-addons} 
\usepackage{graphicx}                    
\usepackage{color}                       
\usepackage{breakurl}                         

\def \apj {Astrophys. J.}

\def \aap {Astron. Astrophys.}

\def \solphys {Solar Phys.}
\def \ssr {Space Sci. Rev}
\def \jgr {J. Geophys. Res}
\def \mdash {$-$}
\begin{document}
\begin{article}
\begin{opening}
\title{Estimation of Arrival Time of Coronal Mass Ejections in the Vicinity
of the Earth Using SOlar and Heliospheric Observatory and Solar TErrestrial RElations Observatory Observations}
\author[addressref={aff1},corref,email={anitha@oa.uj.edu.pl}]{\inits{}\fnm{Anitha}~\lnm{Ravishankar}}
\author[addressref=aff1,email={grzegorz.michalek@uj.edu.pl}]{\inits{}\fnm{Grzegorz}~\lnm{Micha{\l}ek}}

\address[id=aff1]{Astronomical Observatory of Jagiellonian University, Krakow, Poland}

\runningauthor{A.Ravishankar,G.Micha{\l}ek}
\runningtitle{Arrival Time of CMEs}
\begin{abstract}
The arrival time of coronal mass ejections (CMEs) in the vicinity of the Earth is one of the most important parameters in determining space weather. We have used a new approach to predicting this parameter. First, in our study, we have introduced a new definition of the speed of ejection. It can be considered as the maximum speed that the CME achieves during the expansion into the interplanetary medium. Additionally, in our research we have used not only observations from the \emph{SOlar and Heliospheric Observatory} (SOHO) spacecraft but also from  \emph{Solar TErrestrial RElations Observatory} (STEREO) spacecrafts. We focus on halo and partial-halo CMEs during the ascending phase of Solar Cycle 24. During this period the STEREO spacecraft were in quadrature position in relation to the Earth.
We demonstrated that these conditions of the STEREO observations can be crucial for an accurate determination of transit times (TT) of CMEs to the Earth. In our research we defined a new initial velocity of CME, the maximum velocity determined from the velocity profiles obtained from a moving linear fit to five consecutive height--time points. This new approach can be important from the point of view of space weather as the new parameter is highly correlated with the final velocity of ICMEs. It allows one to predict the TTs with the same accuracy as previous models. However, what is more important is the fact that the new approach have radically reduced the maximum TT estimation errors to 29 hours. Previous studies determined the TT with a maximum error equal 50 hours.

\end{abstract}
%
\keywords{Sun: coronal mass ejections (CMEs) . Sun: Space Weather}
\end{opening}
%
 \section{Introduction}
	Coronal mass ejections (CMEs) play an important role in controlling space weather, which can generate the most intensive geomagnetic disturbances on the Earth  (\emph{e.g.},\citealt{Gopalswamy01},\citealt{Gopalswamy02}, \citealt{Gopalswamy002}; \citealt{Srivastava02}; \citealt{Kim}; \citealt{Moon}; \citealt{Manoharan04};  \citealt{Manoharan06}; \citealt{Manoharan10}; \citealt{Manoharan11}; \citealt{Shanmugaraju15}). For geomagnetic-storm forecasting it is crucial to predict when a solar disturbance would reach the Earth. This is not an easy task because the rate of expansion of ejections depends on the magnetic force that drives them and the conditions prevailing in the interplanetary medium. In the initial phase, the magnetic force dominates and the ejection is accelerated rapidly. Farther from the Sun, the propelling force weakens and friction begins to dominate. The ejection speed drops gradually approaching the speed of the solar wind. In addition, the ejection velocity
can change rapidly as a result of CME--CME interactions. Such collisions mostly occur during a maximum of solar activity.

	Initially, models predicting the arrival of interplanetary shocks (IPs) generated by fast CMEs were based on observations of metric Type II radio bursts  (\citealt{Smart85}; \citealt{Smith90}) but these models were inaccurate (\citealt{Gopalswamy98}; \citealt{Gopalswamy001}). \citet{Gopalswamy00} recognized that the distribution of the speed of interplanetary coronal mass ejections (ICMEs) is much narrower (350-650~km~s$^{-1}$) in comparison with the distribution of the speed of CMEs observed near the Sun (150-1050~km~s$^{-1}$). This means that CMEs are effectively accelerated as a result of interaction with the solar wind. During expansion, in the interplanetary medium, their speed gradually approaches the speed of the solar wind. Based on these observations, \citet{Gopalswamy00} introduced an effective acceleration as the difference between the initial [\emph{u}] and final [\emph{v}] speed of an ejection divided by the time [\emph{t}] taken to reach the Earth. They found a definite linear correlation between the effective acceleration [\emph{a}] and initial speed of CMEs: $ a=1.41-0.0035u$ (\emph{a} and \emph{u} are in units of m~s$^{-2}$ and km~s$^{-1}$, respectively). \citet{Gopalswamy000} demonstrated that coronagraphic observations are subject to a major projection effect. To estimate this effect, \citet{Gopalswamy01} used archival data from spacecraft in quadrature (Helios 1 and P78-1). This allowed them to improve the relation between \emph{a} and \emph{u} (\emph{a}=2.193-0.0054\emph{u}).This relation was used to predict the arrival time of CMEs at 1 AU. It was demonstrated that the highest accuracy was obtained when the acceleration ceased at a distance of 0.75 AU. \citet{Michalek04} further developed this approach to predicting the 1 AU arrival time of halo CMEs. They proposed to determine the effective acceleration only from two groups of CMEs, the fastest and slowest events. These events are assumed to not have acceleration cessation at any place between the Sun and Earth. To minimize the projection effect they also used an innovative  method (\citealt{Michalek03}) to obtain the real speed of CMEs. This approach allows one to predict the arrival time of halo CMEs with an average error of 8.7 and 11.2 hours for real and projected initial speeds, respectively.

	Another precaution to be taken in determining the TTs of the CMEs concerns the moment when a given event
reaches the speed of the solar wind (acceleration cessation). However, estimation of the acceleration cessation distance is not an easy task. Few aerodynamic drag models have been developed to solve this problem (\citealt{Vrsnak13}; \citealt{Shanmugaraju14}). These models take into account the difference between speeds of the CMEs and  solar wind. Unfortunately, this approach cannot be applied to all CMEs. Using white-light images and interplanetary scintillations, \citet{Manoharan06} estimated the TTs for 30 CMEs observed from 1998 to 2004. He presented important conclusions showing that the TT can be significantly disturbed by CME--CME interactions and changes in solar-wind properties. Recently \citet{Ibrahim17} have studied 51 halo and partial-halo CMEs in the ascending phase of the Solar Cycle 24 and compared the TT relationship with the initial speed of CMEs in the previous solar cycle, Solar Cycle 23 and the current one, Solar Cycle 24. It has been demonstrated that during the present cycle the CMEs have not been significantly affected by the drag force caused by the interplanetary medium.

	In our current work, we continue this work using a new approach to a more accurately estimate the TT of the CME.
For this purpose, we use images from SOHO \emph{Large Angle and Spectrometric Coronagraphs} (LASCO) and STEREO
\emph{Sun Earth Connection Coronal and Heliospheric Investigation} (SECCHI) coronagraphs and employed a new technique to determine the speed of ejections. After the prolonged minimum of Solar Cycle 23, the ascending phase of Solar Cycle 24 was observed starting from 2009 with an increase in the number of CMEs. At the same time, the STEREO spacecraft achieved 90 degrees separation relative to the Earth, a condition known as quadrature. This location of the spacecrafts allows us to give better definition of the parameters of CMEs, especially of those that are directed towards the Earth (halo CMEs). Additionally, in our studies we introduce a new method for determining the speed of ejections that allows us to estimate the instantaneous speed of CMEs.

	This article is organized as follows. The data and method used for the study are described in Section 2. In Section 3, we present results of our study. Finally, the conclusions and discussions are presented in Section 4.

\section{Data and Method}

	The main aim of the study is to evaluate the TTs of CMEs to the Earth. For this purpose, observations from
the two separate spacecrafts, SOHO/LASCO and STEREO/SECCHI, and a new technique to determine the initial speeds of CMEs were employed.
Since 1995, CMEs have been routinely recorded by the sensitive LASCO (\citealt{Brueckner95}) onboard the SOHO mission.
The SOHO/LASCO instruments had already recorded about 30,000 CMEs by December 2016. The basic attributes of CMEs, determined
manually from running-difference images, among others, are stored in the SOHO/LASCO catalog ({\sf cdaw.gsfc.nasa.gov/CME$\_$list},
\citealt{Yashiro04}, \citealt{Gopalswamy09}). The initial velocity of CMEs obtained by fitting a straight line  to the height--time
measurements has been the basic parameter used in prediction of the TT. This catalog has been widely used for different
scientific studies. Unfortunately, coronagraphic observations of CMEs are subject to projection effects. This makes it practically
impossible to determine the true properties of CMEs and therefore makes it more difficult to forecast their geoeffectiveness.
This effect mostly affects geoeffective events that originate from the disk center.

	Since the launch  of STEREO (\citealt{Kaiser08}) in 2006 we have a unique opportunity to observe the solar corona from
two additional directions. In this study we use these observations to determine velocities of CMEs from additional points of view,
since we are concentrating on the events originating in the ascending phase of the Solar Cycle 24. During this period, the STEREO
spacecraft were approximately in a quadrature configuration with respect to the Earth. Using quadrature observations with the two
STEREO spacecrafts, we can estimate the  plane-of-sky speeds which is close to the true radial speed of events ejected from the
disk center. This was demonstrated by \citet{Bronarska18}. To obtain the STEREO speeds of CMEs we have performed
identical manual measurements as in the case of the LASCO observations (\citealt{Yashiro14}). The only difference was that for
these measurements we used COR2 coronagraphs and the optical telescopes: HI1. To determine the speed of a given CME, we employed
only images from the STEREO-A or-B spacecrafts, which showed a better quality of observation. This approach allows us to obtain
the most accurate height--time data points. As was demonstrated by \citet{Michalek17} the maximal errors in estimation of velocity
significantly depend on the quality of CMEs recorded by LASCO coronagraphs. They also demonstrated that a number of height--time points
measured for a particular event is the dominant factor in determining the accuracy of CME parameters to the greatest extent.
This number is directly dependent on the quality of observations, instrument data gaps, and CME speed. We must also mention that
in the case of our considerations it is not important which STEREO spacecrafts (A or B) observes a given event because our study
is carried out in the period when these twin spacecrafts were in quadrature in relation to the Earth. In addition, these studies
are concentrated on halo CMEs that are formed in the central part of the solar disk and are directed towards the Earth. Having the
STEREO height--time measurements, we could obtain initial speeds of CMEs from a linear fit recorded by instruments onboard the
STEREO spacecrafts. These speeds have been calculated in an identical manner, with the exception of the instruments used, as in the
case of those included in the SOHO/LASCO catalog.

	It is worth emphasizing here why STEREO and SOHO were used to derive speeds separately. In the current research, we focus
on halo CMEs. The STEREO observations in quadrature provide speeds that are very close to spatial (real) velocities, whereas
measurements with SOHO provide speeds that are significantly modified by projection effects. We are interested in how these
different speeds can be used to determine the TT of a CME to the Earth.

	Until now, empirical models predicting the TTs of the CMEs have employed the initial velocities of ejections
obtained from a linear fit to all manually measured height--time data points. Therefore, the determined speeds are in some sense
 the average velocities of CMEs in the field of view of the respective coronagraphs. It is obvious that the instantaneous  velocities
 of CMEs change with distance from the Sun, since initially their speeds increase when their dynamics are dominated by the Lorentz
 forces, and they reach their maximum speed when the Lorentz force balances the friction force. From this moment, the CMEs are
 slowed down until they reach the speed of the solar wind. Therefore, it is evident that the speed determined from the linear
 fit depends not only on the actual CME speed but also on the number of data points and this significantly depends on the brightness
 of a given ejection.

	In this context, it is worthwhile to estimate the CME speed using a different approach, i.e. to determine the initial speed
 of CMEs based on their maximal velocities. For this approach, in our current work, we employed a simple technique to determine
  instantaneous  velocities of the CMEs.  To obtain these velocities we also used linear fits to height--time points but for now we
 used limited number of these points. In our study we considered linear fits using three to eight height--time data points only.
 Shifting successively, in this way we can obtain the instantaneous speed of the CME. Using such a linear fit, for all of the
 height--time data points measured for a given CME, we obtained instantaneous profiles of velocities in time or in distance
 from the Sun.

	From the velocity profiles thus obtained, we can easily determine the maximal velocity, as well as the time and the distance when this speed has been achieved. We tested this method employing different number of height--time points (from three to eight). The most
reasonable results were obtained when we employed five successive height--time points for a linear fit, hence we used this method
in our present study. Formally, two neighboring height--time points are enough to calculate the instantaneous speed.
Unfortunately manual measurements are subject to unpredictable random errors. These errors result from the subjective nature of
manual measurements.

	In order to minimize the impact of these errors on the determined instantaneous speed we decided to apply linear fits.
This technique allows us to obtain smooth profiles of instantaneous speed. Applying this method we have determined profiles of
instantaneous velocities of CMEs in the field of view of LASCO (C2 and C3) and STEREO telescopes. In the case of the STEREO twin
spacecraft, we used only observations from the one in which the quality of observation was better. We employed subscript $_5$ to denote that we have used five successive height--time points for linear fit to obtain the maximal velocity. These profiles of instantaneous
velocities can be easily used to determine the  maximal/\emph{V}$_5$ velocity in SOHO (\emph{V}$_{5-\mathrm{SOHO}}$) and STEREO
(\emph{V}$_{5-\mathrm{STEREO}}$) telescopes.  It should be mentioned that these velocities are the plane-of-sky speeds. However,
\emph{V}$_{5-STEREO}$ for halo CMEs is very close to the true radial speed. Having the profiles of instantaneous velocities we estimated
the time [T$_{5-\mathrm{SOHO}}$ and T$_{5-\mathrm{STEREO}}$] and distance [D$_{5-\mathrm{SOHO}}$ and D$_{5-\mathrm{STEREO}}$] when CMEs reached the maximal speeds [\emph{V}$_{5-\mathrm{SOHO}}$] and STEREO [\emph{V}$_{5-\mathrm{STEREO}}$] for the respective coronagraphs.

	It is obvious that CMEs, when moving from the Sun to the Earth, are subject to three different phases of propagation.
First, close to the Sun, they are subject to rapid initial acceleration [phase 1, S1]. At the end of this phase they reach their
maximum speeds [\emph{V}$_{5-\mathrm{SOHO}}$] or STEREO [\emph{V}$_{5-\mathrm{STEREO}}$]. Then, when their dynamics is determined by the drag force (due to interaction with solar wind), they move with a negative acceleration [phase 2, S2]. After reaching the speed of the solar
wind, they move at a constant speed ([\emph{V}$_\mathrm{CONST}$], the average speed of the solar wind) until they reach Earth's orbit
[phase 3, S3]. Using our definitions we may write:
\begin{equation}
S1+S2+S3=1 \mathrm{AU} \label{au}
\end{equation}
\begin{equation}
S1=D_{5-\mathrm{SOHO}}/D_{5-\mathrm{STEREO}},  \label{S1}
\end{equation}
where S1, S2, and S3 are the distances that the CME travels in the next three phases of propagation.
In our work, we focus only on the last two phases of propagation. During these phases of propagation a given CME traverses the
respective distances: we have
\begin{equation}
S2=V_{\mathrm{MAX}}T2-{aT2^{2}\over 2}   \label{vmax}
\end{equation}
and
\begin{equation}
S3=V_{\mathrm{CONST}}T3,  \label{s3}
\end{equation}
where $V_{\mathrm{MAX}}$ is the maximal velocity of CMEs achieved in the respective telescope [\emph{V}$_{5-\mathrm{SOHO}}$] or
[\emph{V}$_{5-\mathrm{STEREO}}$], T2 is the travel time during the second phase of propagation, $a$ is acceleration during the second
phase of propagation, \emph{V}$_{\mathrm{CONST}}$ is the velocity of propagation during the third phase of propagation and T3 is the
travel time during the third phase of propagation. These equations form the basis of our further considerations, especially
those relating to the TTs of CMEs from the Sun to the Earth. In this approach the TT of a CME is $T2+T3$
and the distance to travel is $S1+S2$. These equations fully describe the kinematics of CMEs in the field of view of LASCO and
SECCHI coronagraphs. The only undefined parameter is the acceleration [$a$] of CMEs in the second phase of their propagation.
Below, we present various methods for its determination. This allows us to test different models used to determine the TTs of CMEs from the Sun to the Earth.

In the paper we consider different methods to
determine velocities of CMEs. These velocities can be correlated with the TTs.
Fitting curves to TT--velocity points we built theoretical models that can be
used to predict the TT in the future. For individual CMEs, we can determine
the error in estimation of TT as the difference between the TT determined on the
basis of the model and the actual observations. Having these errors for a
given model and entire populations of considered CMEs we can determine the
average absolute and maximal errors. This means that the maximal error for
a given model is the maximal error from the distribution.

 \section{Results}

	Our study concentrates on the ascending phase of the Solar Cycle 24. During this period we were able to record,
at the same time, CMEs observed by STEREO-A and -B spacecrafts as they were separated by 90 degrees with respect to Earth.
\citet{Ibrahim17} compiled a list of halo and partial-halo CMEs during 2009\,\--2013. Among them, they were able to identify
the ICME at the Earth's vicinity for 51 events. These events are the basis of our study. Their analysis were limited to the data
included in the SOHO/LASCO catalog. Additionally, in our research, for each CME included in their list, we conducted the analysis
which was described in the previous section. This means that for each CME we have measured height--time points in the STEREO field
of view and the initial ejection velocities of CMEs [\emph{V}$_{\mathrm{AVG-STEREO}}$ and \emph{V}$_{5-\mathrm{STEREO}}$]  were determined.
A few events were too faint in the STEREO images so we were not able to obtain height--time points for them. Having the profiles
of instantaneous velocities, we estimated the time and distance when the CMEs reached their maximal speeds. All of these data are
shown in Table 1. The near-Sun observational details in the LASCO field of view are given in columns two -- seven. The onset date and
time of CME ejection are in columns two and three, respectively. The average velocity from a linear fit to all data points
(from the SOHO/LASCO catalog), the maximal velocity from linear fit to five successive height--time points, distance,
and the time when a given CME reaches the maximal velocity are displayed in column four -- seven, respectively. Next, the TTs
of ICMEs and shocks and final velocity of ICME in the vicinity of Earth received by the $in-situ$ observations made by $Advanced
Composition Explorer$ (ACE) instruments are given in columns eight -- ten, respectively. These data are from
\citet{Ibrahim17}. The details from observations in the STEREO field of view are shown in columns eleven -- fourteen.
In the respective columns we find the average velocity from a linear fit to all data points, the maximal velocity from
linear fit to five successive height--time points, distance, and time when a given CME reaches the maximal velocity.
The data shown in the table are the basis for our calculation of the TT of CME from the Sun to the Earth.
The results are presented in the following sections. In the table are included all CMEs (51) considered by
\citet{Ibrahim17} having recognised magnetic cloud structure at the Earth (having determined the TT for an ICME).
For 48 and 39 (including interacting events) of them we were able to obtain the maximal velocities (\emph{V}$_{5-\mathrm{SOHO/STEREO}}$)
in the SOHO/LASCO and STEREO field of view, respectively.

%
\begin{sidewaystable}
\begin{tiny}

\centering
\caption{Observational parameters of 51 ICMEs in the period 2009-2013}
\begin{longtable}{|c|c|c|c|c|c|c|c|c|c|c|c|c|c|c|c|}

\hline
\multirow{3}{*}{\#}&\multicolumn{7}{|c|}{SOHO CME
Observations}&\multicolumn{3}{|c|}{In-situ Observations}&\multicolumn{5}{|c|}{STEREO
Observations}\\
&Date & Time & \emph{V}$_{\mathrm{AVG}}$& \emph{V}$_5$& R$_{\mathrm{MX}}$&        T$_{\mathrm{MX}}$& TT$_{\mathrm{MX}}$ & TT$_{\mathrm{ICME}}$&
TT$_{\mathrm{SHO}}$&\emph{V}$_{\mathrm{FIN}}$&\emph{V}$_{\mathrm{AVG}}$& \emph{V}$_5$& R$_{\mathrm{MX}}$ & T$_{\mathrm{MX}}$ & TT$_{\mathrm{MX}}$\\
& DD/MM/YYYY & [HH:MM] & [km\,s$^{-1}$] & [km\,s$^{-1}$] & [R$_{\mathrm{\odot}}$] & [HH:MM] & [HH] & [HH:MM] & [HH:MM] &
[km\,s$^{-1}$] & [km\,s$^{-1}$] & [km\,s$^{-1}$] & [R$_{\mathrm{\odot}}$] & [HH:MM] & [HH] \\
\hline
1 & 16/12/2009 & 04:30 & 276 & 328 & 10 & 10:44 & 132 & 138:29 & 134:17 & 365 & 416
& 511 & 14 & 09:09 & 134 \\

2 & 07/02/2010 & 03:54 & 421 & 443 & 13 & 08:44 & 94 &  99:15 &  93:02 & 364 & 418 &
680 & 15 & 07:12 & 96  \\

3 & 12/02/2010 & 13:42 & 509 & 577 & 14 &15:09 & 77 &  78:28 &  77:04 & 317 & 637 &
789 & 14 & 14:57 &  72 \\

4 & 12/02/2010 & 22:30 & 1180 & 1387 & 11 & 23:49 & 68 &  69:16 &  68:16 & 317 & 572
& 767 & 14 &  01:22 &  68 \\

5 & 08/04/2010 & 04:54 & 264 & 315 & 09 &  05:30 &  88 &  89:00 &  79:06 & 423 & 534
& 698 & 55 & 06:13 &  87 \\

6 & 23/05/2010 & 18:06 & 258 & 340 & 14 & 02:44 & 113 &  122:07 &  104:42 & 383 &
419 & 669 & 13 & 22:04 &  118 \\

7 & 24/05/2010 & 14:06 & 427 & 490 & 17 &20:44 & 95 &  102:07 &  84:42 & 383 & 636 &
1056 & 14 &  18:07 &  98 \\

8 & 22/06/2010 & 10:50 & 167 & 218 & 07 & 17:44 & 87 &  94:44 &  87:45 & 452 & -- &
-- & -- &  -- &  -- \\

9 & 15/02/2011 & 02:24 & 669 & 837 & 08 & 03:42 & 88 &  90:15 &  71:14 & 538 & 611 &
1097 & 06 &  02:38 &  90 \\

10 & 01/06/2011 & 18:36 & 361 & 507 & 08 &  21:54 & 75 &  78:18 &  74:08 & 520 & 425
& 740 & 80 &  03:45 & 78 \\

11 & 02/06/2011 & 08:12 & 976 & 1180 & 19 &  11:18 & 62 &  65:06 &  60:32 & 520 &
650 & 1324 & 15 &  10:07 & 63 \\

12 & 21/06/2011 & 03:16 & 719 & 835 & 15 &  04:01 & 50 &  51:15 &  47:09 & 591 & 844
& 1058 & 25 &  06:49 & 47 \\

13 & 02/08/2011 & 06:36 & 712 & 936 & 10 &  08:06 & 67 &  69:35 &  63:23 & 404 & 808
& 1159 & 16 &  08:42 & 67 \\

14 & 03/08/2011 & 14:00 & 610 & 841 & 05 &  14:20 & 64 &  64:32 &  51:41 & 549 &
1043 & 1598 & 15 &  15:24 & 63 \\

15 & 06/09/2011 & 02:24 & 575 & -- & -- & -- & -- &  93:00 &  82:00 & 350 & 695 &
1246 & 15 &  00:51 & 71 \\

16 & 07/09/2011 & 23:05 & 792 & 916 & 05 &  23:21 & 71 &  71:41 &  61:05 & 319 & 642
& 810 & 04 &  23:08 & 71 \\

17 & 01/10/2011 & 09:36 & 448 & 620 & 08 &  11:06 & 95 &  97:02 &  94:02 & 558 &
361 & 737 & 15 &  02:42 & 79 \\

18 & 26/12/2011 & 11:48 & 736 & 899 & 04 &  12:08 & 70 &  70:26 &  46:16 & 461 & 555
& 906 & 14 &  14:22 & 68 \\

19 & 23/01/2012 & 04:00 & 2175 & 2285 & 23 &  05:42 & 41 &  43:35 &  34:57 & 350 &
1347 & 2288 & 14 &  04:51 & 42 \\

20 & 19/02/2012 & 20:57 & 539 & 736 & 05 &  21:40 & 73 &  74:20 &  53:32 & 577 & --
& -- & -- &  -- & -- \\

21 & 07/03/2012 & 00:24 & 2684 & 2789 & 09 &  00:49 & 52 &  52:17 &  34:18 & 500 &
974 & 1743 & 07 &  00:53 & 52 \\

22 & 07/03/2012 & 01:30 & 1825 & 1946 & 10 &  01:54 & 51 &  51:23 &  33:24 & 717 &
1151 & 1552 & 07 &  01:38 & 51 \\

23 & 10/03/2012 & 18:00 & 1296 & 1392 & 12 & 19:18 & 50 &  51:43 &  39:21 & 704 &
1238 & 2012 & 14 &  18:59 & 50 \\

24 & 26/03/2012 & 23:12 & 1390 & 1565 & 07 & 23:31 & 31 &  32:12 &  24:12 & 424 & --
& -- & -- &  -- & -- \\

25 & 28/03/2012 & 01:36 & 1033 & 1326 & 05 & 02:25 & 61 &  62:22 &  50:23 & 395 & --
& -- & -- &  -- & -- \\

26 & 18/04/2012 & 09:24 & 448 & 669 & 15 & 14:06 & 124 &  128:06 &  113:00 & 384 &
-- & -- & -- &  -- & -- \\

27 & 18/04/2012 & 17:24 & 581 & 763 & 06 &  18:16 & 120 &  120:06 &  105:00 & 384 &
407 & 1130 & 08 &  18:38 & 119 \\

28 & 14/06/2012 & 14:12 & 987 & 1151 & 18 &  16:54 & 53 &  56:34 &  43:06 & 515 &
738 & 1296 & 05 &  14:23 & 56 \\

29 & 28/06/2012 & 20:00 & 1313 & 1570 & 07 &  20:25 & 44 &  44:34 &  25:29 & 660 &
-- & -- & -- &  -- & -- \\

30 & 02/07/2012 & 06:24 & 988 & 1186 & 11 &  08:06 & 63 &  65:20 &  58:23 & 503 &
722 & 842 & 25 &  12:49 & 59 \\

31 & 02/07/2012 & 08:36 & 1074 & 1251 & 12 &  09:54 & 61 &  63:08 &  56:35 & 503 &
-- & -- & -- &  -- & -- \\

32 & 12/07/2012 & 16:48 & 885 & -- & -- &  17:12 & -- &  62:00 &  48:00 & 600 & 664
& 1351 & 07  &  17:23 & 61\\

33 & 21/11/2012 & 16:00 & 529 & 807 & 06 &  16:49 & 67 &  80:11 &  65:58 & 410 & 598
& 799 & 15 &  19:34 & 64 \\

34 & 23/11/2012 & 13:48 & 519 & 680 & 08 &  15:49 & 68 &  70:18 &  62:11 & 517 & 554
& 807 & 16 &  18:00 & 65 \\

35 & 23/11/2012 & 23:24 & 1186 & 1450 & 06 &  23:45 & 60 &  60:32 &  53:13 & 517 &
-- & -- & -- &  -- & -- \\

36 & 31/01/2013 & 06:36 & 682 & 925 & 23 &  12:18 & 75 &  80:05 &  73:23 & 444 & 750
& 1211 & 15 &  09:39 & 77 \\

37 & 06/02/2013 & 00:24 & 1867 & 1923 & 18 &01:37 & 55 &  56:15 &  52:01 & 419 &
1084 & 1652 & 17 & 02:33 & 54 \\

38 & 27/02/2013 & 04:00 & 622 & 928 & 19 &  10:06 & 53 &  59:13 &  50:46 & 532 & 481
& 682 & 12 &  08:55 & 54 \\

39 & 15/03/2013 & 07:12 & 1063 & 1388 & 18 & 09:54 & 53 &  55:06 &  46:40 & 650 &
669 & 1382 & 14 &  08:51 & 54 \\

40 & 11/04/2013 & 07:24 & 861 & 1099 & 05 & 07:45 & 81 &  82:13 &  63:32 & 457 & 677
& 1153 & 13 & 09:26 & 80 \\

41 & 21/04/2013 & 07:24 & 919 & -- & -- & -- & -- &  71:00 &  59:00 & 332 & 661 &
949 & 15 &  11:21 & 67 \\

42 & 21/04/2013 & 16:00 & 857 & 1113 & 10 &  17:30 & 61 &  63:21 &  50:28 & 332 &
635 & 879 & 05 &  16:53 & 62 \\

43 & 14/05/2013 & 23:12 & 667 & 1152 & 05 &  23:24 & 87 &  87:07 &  73:19 & 418 &
736 & 1049 & 14 &  02:54 & 84 \\

44 & 22/06/2013 & 18:24 & 477 & 710 & 06 &  19:37 & 128 &  141:18 &  128:17 & 413 &
-- & -- & -- &  -- & -- \\

45 & 28/06/2013 & 02:00 & 1037 & 1289 & 22 & 02:25 & 48 &  49:04 &  32:37 & 473 &
678 & 1429 & 08 & 02:38 & 48 \\

46 & 24/09/2013 & 20:36 & 919 & 1015 & 05 & 00:27 & 46 &  66:03 &  53:28 & 324 & --
& -- & -- & -- & -- \\

47 & 06/10/2013 & 14:43 & 567 & 705 & 10 & 17:56 & 62 &  120:03 &  116:13 & 346 & 339
& 1232 & 11 & 17:50 & 62 \\

48 & 16/10/2013 & 15:48 & 514 & 623 & 05 & 16:32 & 119 &  120:44 &  116:15 & 304 &
-- & -- & -- & -- & -- \\

49 & 24/10/2013 & 01:25 & 399 & 600 & 04 & 01:45 & 113 &  133:28 &  129:23 & 321 &
292 & 448 & 12 & 06:32 & 131 \\

50 & 07/11/2013 & 10:36 & 1405 & 1712 & 08 &  10:55 & 41 &  53:31 &  33:03 & 603 &
-- & -- & -- &  -- & -- \\

51 & 28/12/2013 & 17:36 & 1118 & 1247 & 13 &  17:52 & 55 &  55:26 &  54:15 & 423 &
778 & 1293 & 13 &  19:26 & 53 \\
\hline

\end{longtable}
\end{tiny}
\end{sidewaystable}

   \subsection{Velocities of CMEs}
   \begin{figure}[h]
\centerline{\includegraphics[width=1.0\textwidth,clip=]{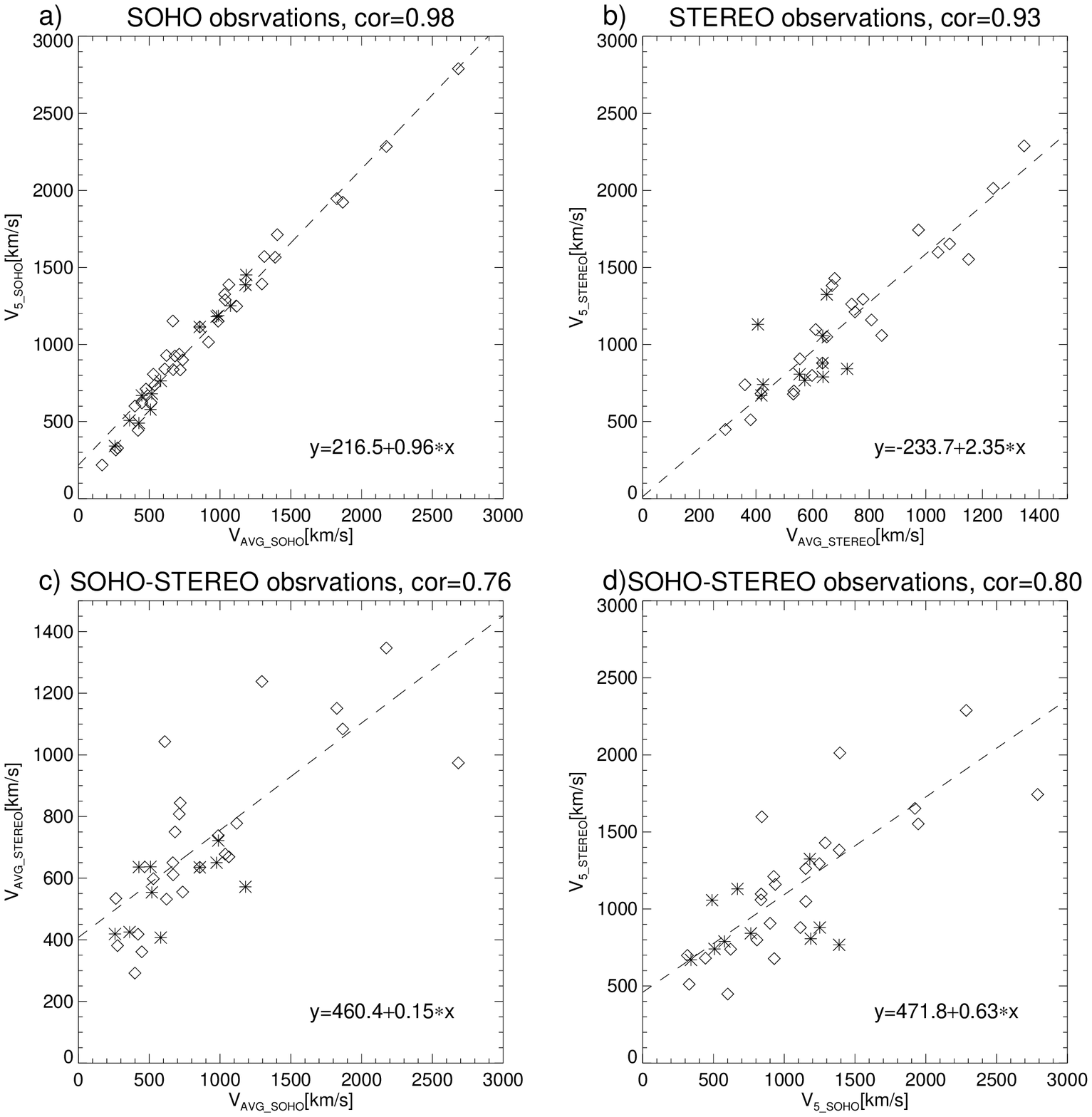}}
 \caption{Relationship between (\textbf{a}) the average and maximal/\emph{V}$_5$ speed determined in SOHO images. (\textbf{b}) the average and maximal/\emph{V}$_5$  speed determined in the STEREO images. (\textbf{c}) the average speed determined in SOHO and STEREO images. (\textbf{d}) the maximal/\emph{V}$_5$  speed determined in SOHO and STEREO images. \emph{Dashed lines} are linear fits to data points. $Formulas$ representing these linear fits are placed in the \emph{lower-right corners of each panel}. \emph{Open diamond symbols} are for non-interacting and \emph{star symbols} are for interacting CME, respectively.}
 \end{figure}

	In the previous sections, we presented methods for determining the initial speeds of ejections.
In this section we present the relationship between these speeds obtained from SOHO and STEREO images. In Figure 1,
the successive panels show the relationships between the average and maximal/\emph{V}$_5$ speeds determined in SOHO images,
the average and maximal/\emph{V}$_5$ speeds determined in the STEREO images, the average speeds determined in SOHO and STEREO
images and the maximal/\emph{V}$_5$ speeds determined in SOHO and STEREO images. Dashed lines are linear fits to data points.
Formulas representing linear fits are placed in the lower-right corners of each panel. It can be seen  that the average ejection
velocities are strongly correlated with their maximal velocities (Panels a and b), regardless of the instrument used for
their determination.  We can notice that the maximal velocities are much larger (on average 80$\,\%$) than the average velocities
in the case of observations from the STEREO spacecrafts (Panel b). For SOHO observations the maximal velocities are on average
only 25$\,\%$ larger than the average velocities. This results from the fact that the field of view of STEREO instruments used
for determining velocity profiles (COR2 and HI1) is much larger than the field of view of LASCO coronagraphs (C2 + C3).
It means that the field of view of the STEREO telescopes covers the area where CMEs undergo significant deceleration due to
interaction  with the solar wind. For this reason, the average velocities of the CMEs determined from the STEREO observations
are significantly lower than other speeds determined in these studies.
	
	Correlations between the speeds for these two instruments are slightly smaller. The correlation coefficients
are 0.75 and 0.80 respectively for the average and maximal speeds. In this case, the larger dispersion of speeds results
from the fact that they are determined from the two different instruments (SOHO and STEREO) that observe the Sun at different
angles. Depending on the source location on the solar disk and the position of the spacecraft, the determined speeds are
subject to different projection effects (\citealt{Bronarska18}). This effect, among others, is the reason that the determined
speeds may be different for each of the telescopes. In our considerations \emph{V}$_{5-\mathrm{SOHO}}$ are on average smaller
(about 50~km~s$^{-1}$) in comparison with \emph{V}$_{5-\mathrm{STEREO}}$. This is due to projection effects. Earth-directed CMEs recorded in SOHO/LASCO coronagraphs are subject to more significant projection effects in comparison to observations by
the STEREO spacecrafts in a quadrature position.

	In Figure~2, relationships between the initial speeds and speeds of interplanetary coronal mass ejections (ICMEs)
obtained from $in-situ$ measurements are shown. In successive panels we have displayed scatter plots of the average velocities
and the maximal/\emph{V}$_5$  velocities determined by SOHO and LASCO telescopes $versus$ ICME speeds recorded near the Earth.

\begin{figure}
 \centerline{\includegraphics[width=1.0\textwidth,clip=]{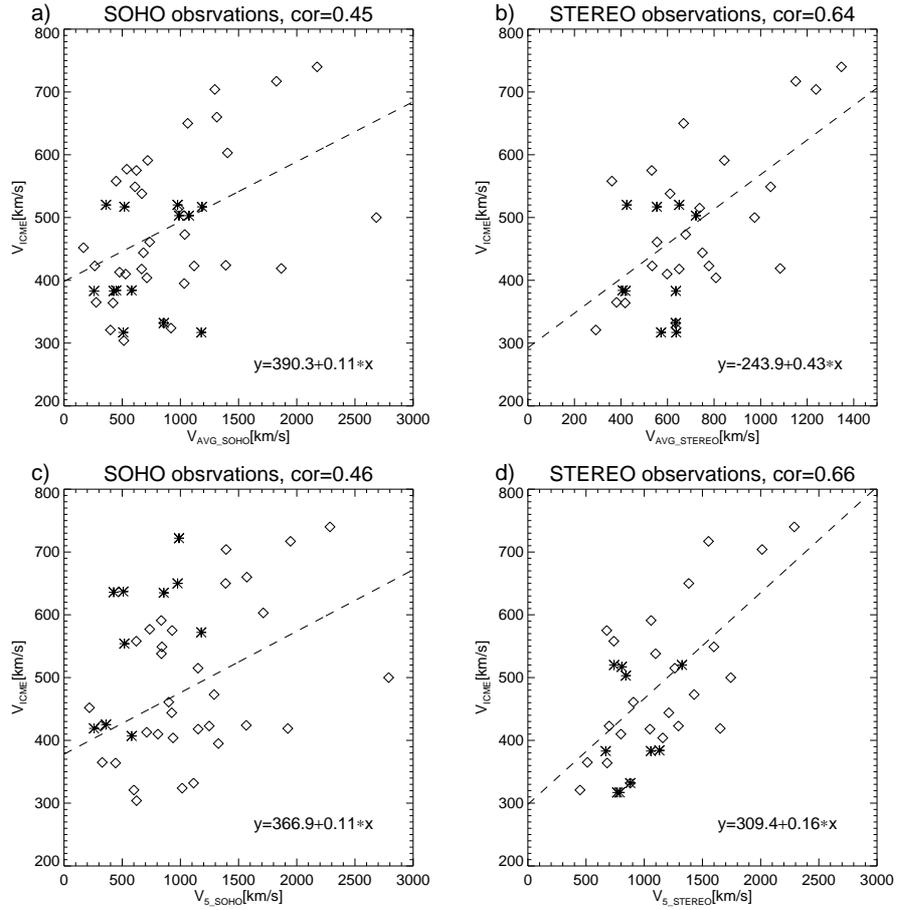}}
 \caption{The scatter plots of the average and maximal/\emph{V}$_5$  speeds determined in SOHO and STEREO images versus ICME speeds determined near the Earth.  \emph{Dashed lines} are linear fits to data points. $Formulas$ presenting these linear fits are placed in the \emph{lower-left corner of each panel}. \emph{Open diamond symbols} are for non-interacting and \emph{star symbols} are for interacting CMEs, respectively.}
 \end{figure}

	For the initial speed determined in the SOHO images, the correlation coefficients are less than 0.5. However,
in the case of the STEREO spacecraft these correlation coefficients are significantly higher (up to 0.66). It is worth noting
that the the most significant correlation is between the final speeds of ICMEs and maximal/\emph{V}$_5$ speeds obtained from
the STEREO images (Panel d). From the point of view of space weather this is a new and very important result. The speed of an
ICME is one of the most important parameters determining the geoeffectivness of CMEs. These relations for speeds obtained from
the STEREO telescopes allow for a more precise estimation of ICME velocities in the vicinity of the Earth and thus the
prediction of their impact on the Earth becomes more accurate.

\subsection{Velocities and Transit Time to the Earth}

	Depending on the velocity used, in our current research, we define the TT of a CME in two ways.
For the average velocities determined in the STEREO or SOHO images, the TT is the time difference between the CME onset time in
LASCO-C2 field of view [T$_{\mathrm{CME}}$] and the ICME arrival time [T$_{\mathrm{ICME}}$] at the Earth's vicinity using $in-situ$ observations
[TT=T$_{\mathrm{CME}}$-T$_{\mathrm{ICME}}$]. In the case of maximal velocities obtained in the STEREO or SOHO images, the TT is the difference
between the time when the CME reaches its maximal speed [T$_{\mathrm{MAX}}$] and the ICME arrival time [T$_{\mathrm{ICME}}$] at the Earth's
vicinity using $in-situ$ observations [TT=T$_{\mathrm{MAX}}$-T$_{\mathrm{ICME}}$]. The TTs for shock generated by ICMEs are obtained in similar ways.
The relationships between TTs and CME speeds are shown in the following figures.

\begin{figure}
 \centerline{\includegraphics[width=1.0\textwidth,clip=]{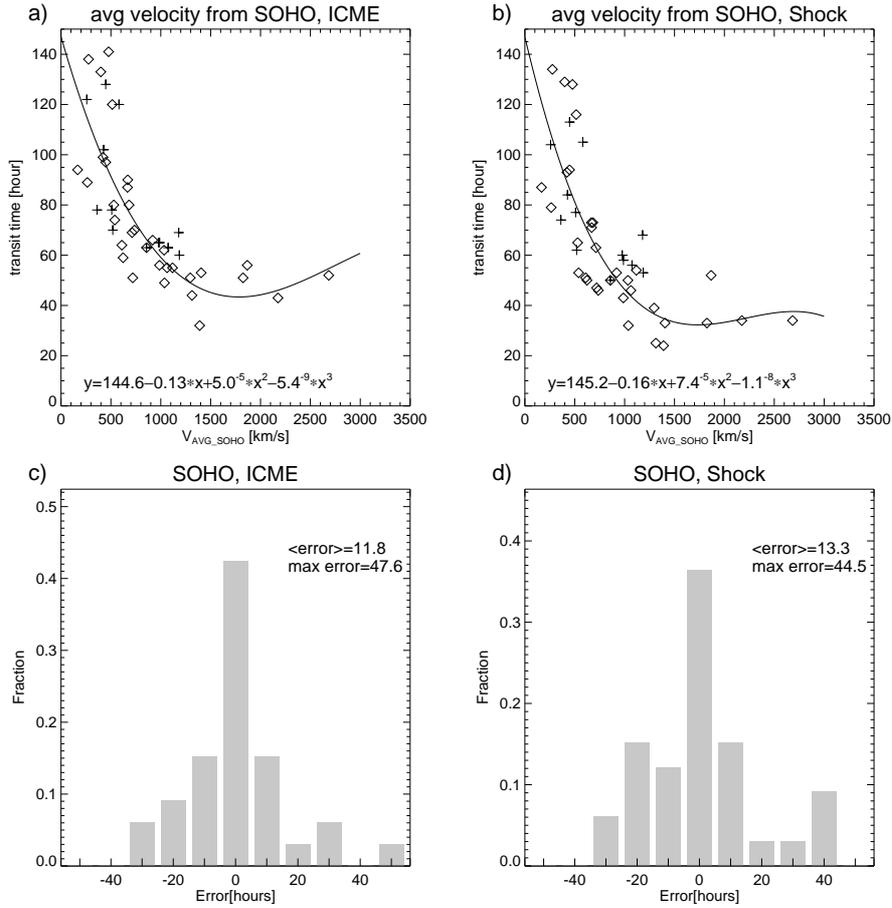}}
 \caption{Relationship between the average CME speed obtained in the SOHO images and the observed TT for ICME (Panel \textbf{a}) and IP shock (Panel \textbf{b}). \emph{Open diamond symbols} are for non-interacting and \emph{star symbols} are for interacting CMEs, respectively.
The third-order polynomial relationship between the observed TT and the speed is indicated by a \emph{dashed curve}. The distributions of errors between predicted and observed TTs are presented in bottom panels (Panel \textbf{c} for ICME and Panel \textbf{d} for IP shock). The values of average and maximal errors are presented in the panels.}
 \end{figure}

	As shown in the Figure~3, the TTs for CME with speeds below 1000~km~s$^{-1}$ are in the range 50-140 hours.
The TTs for fast events (\emph{V}$>$1000~km~s$^{-1}$) are in the range 40-60 hours. The third-order polynomial relationship
between the observed TT and the speed is indicated by a dashed curve. This fitting can be considered as an empirical model
to predict the TT. The difference between the observed and predicted TTs for a given CME speed can be considered as an error
by the model. The lower panels show the distribution of errors in determining the TT for this empirical model. 
The average errors are about $\approx$ 11-12~hours and the maximum errors are very significant and reach values $\approx$50~hours. These errors are the maximum of the distribution of errors. Maximum errors were determined as the maximum difference in time between the theoretical model and the observational data. These errors determined only for non-interacting CMEs.

 \begin{figure}
 \centerline{\includegraphics[width=1.0\textwidth,clip=]{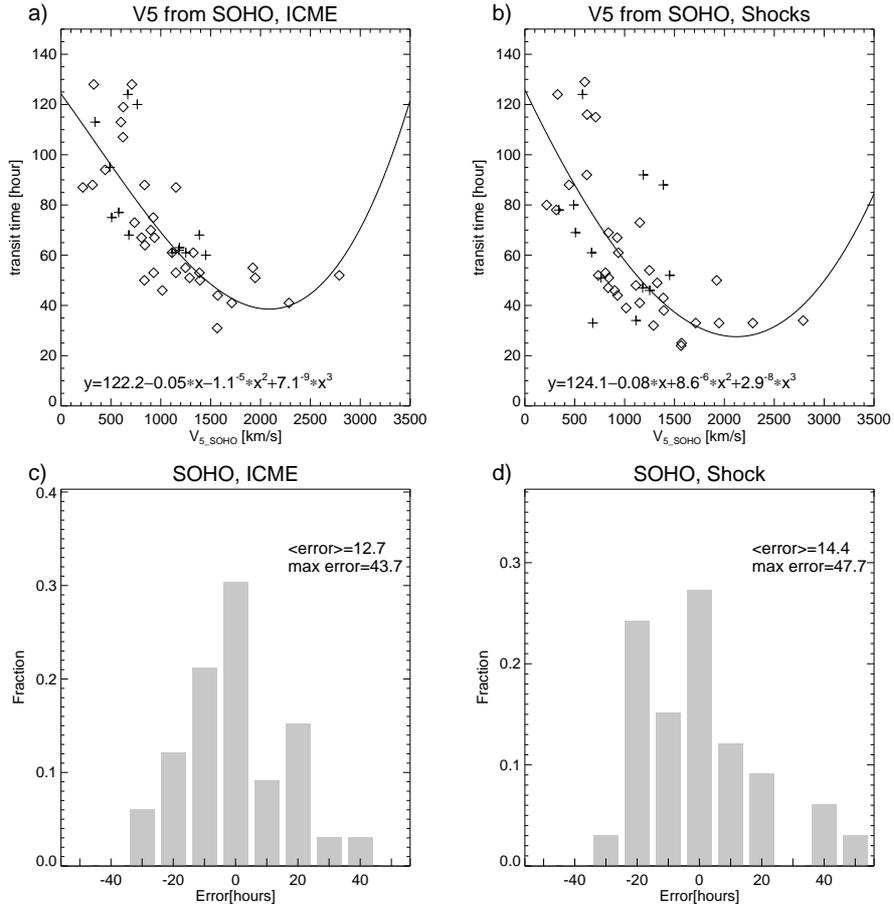}}
 \caption{Relationship between the maximal/\emph{V}$_5$ CME speed  and the observed TT for ICME (Panel \textbf{a}) and IP shock (Panel \textbf{b}). \emph{Open diamond} symbols are for non-interacting and \emph{star symbols} are for interacting CMEs, respectively. The third-order polynomial relationship between the observed TT and the speed is indicated by a \emph{dashed curve}. The distributions of errors between predicted and observed TTs are presented in bottom panels (Panel \textbf{c} for ICME and Panel \textbf{d} for IP shock). The values of average and maximal errors are presented in the panels.}
 \end{figure}

	In Figure~4 we display the same data but for the maximal velocity/\emph{V}$_5$ obtained from the SOHO images.
As shown in the figure, results are similar but the average errors are slightly higher, $\approx$13 hours. This means that the
introduction of the maximal velocity of CME has no effect on a more accurate TT prediction. Figures~3~and~4 and the above
discussion refer to SOHO observations.

 \begin{figure}
 \centerline{\includegraphics[width=1.0\textwidth,clip=]{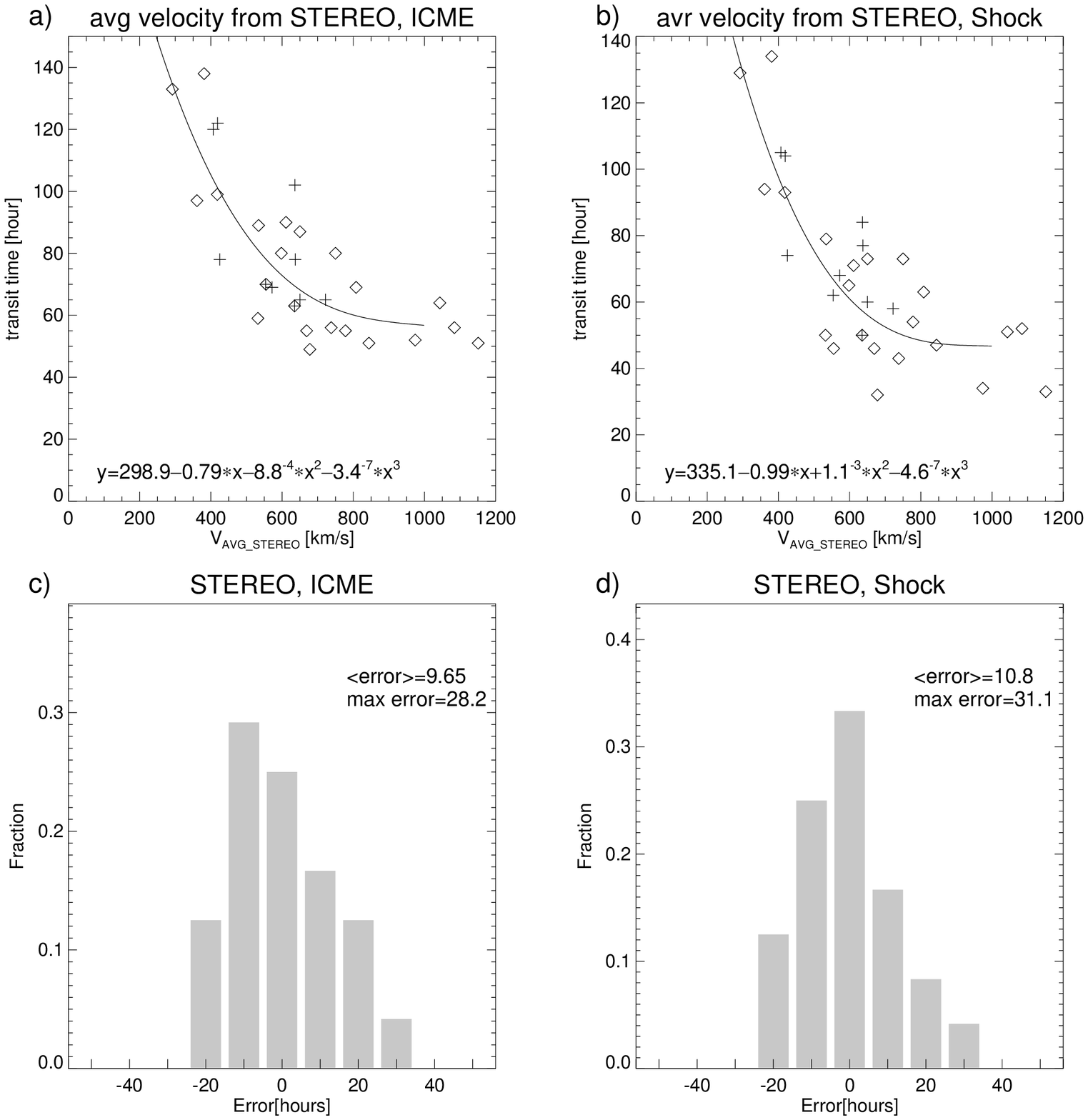}}
 \caption{Relationship between the average CME speed obtained from the STEREO images and the observed TT for ICME (Panel \textbf{a}) and IP shock (Panel \textbf{b}). \emph{Open diamond symbols} are for non-interacting and \emph{star symbols} for interacting CMEs, respectively. The third-order polynomial relationship between the observed TT and the speed is indicated by a \emph{dashed curve}. The distributions of errors between predicted and observed TT are presented in bottom panels (Panel \textbf{c} for ICME and Panel \textbf{d} for IP shock). The values of average and maximal errors are presented in the panels.}
 \end{figure}

	In Figure~5, results for the average speeds determined in the STEREO images are presented. As shown in the figure,
the TT is significantly related to the average speed. The data points are not scattered around the empirical model represented
by a third-order polynomial fit. This fit is represented by  a dashed curve. This empirical model can be used, with great accuracy,
to predict TTs of CMEs. In this case the average errors in the prediction of the TT are only $\approx$9 and 10 hours for ICMEs
and IP shock, respectively. However, a more significant fact is that in this case the maximal errors are much lower than in the previously presented models (\citealt{Gopalswamy01}, \citealt{Michalek04}; \citealt{Manoharan06}), i.e. 29 and 38 hours
 for ICME and IP shock, respectively. Similar results were obtained for the maximal/\emph{V}$_5$ speed in the STEREO images.
Results for these considerations are shown in Figure~6.

 \begin{figure}
 \centerline{\includegraphics[width=1.0\textwidth,clip=]{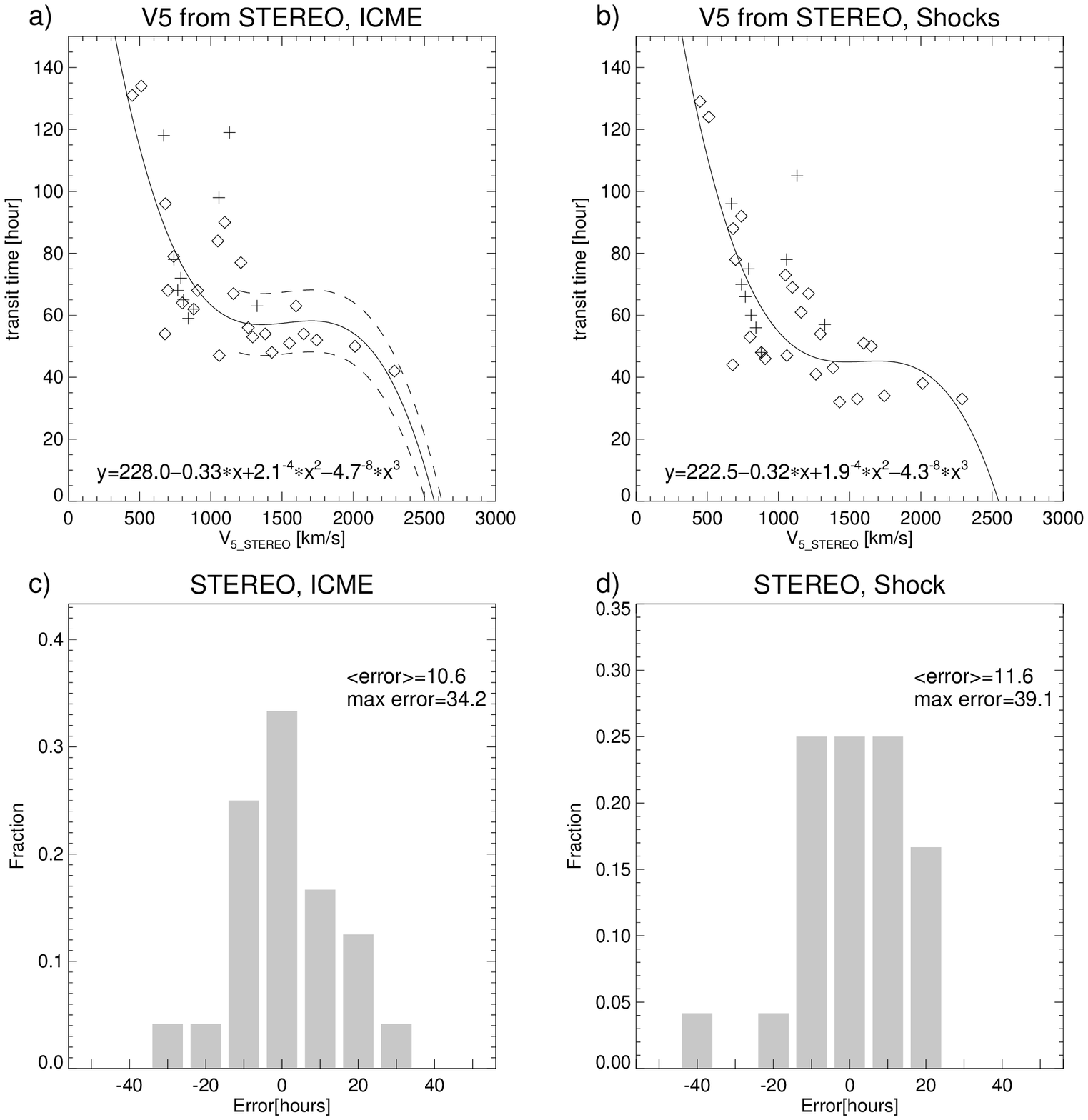}}
 \caption{Relationship between the maximal/\emph{V}$_5$ CME speed obtained from the STEREO images and the observed TT for ICME (Panel \textbf{a}) and IP shock (Panel \textbf{b}). \emph{Open diamond} symbols are for non-interacting and \emph{star symbols} are for interacting CMEs, respectively. The third-order polynomial relationship between the observed TT and the speed is indicated by a \emph{dashed curve}. The distributions of errors between predicted and observed TTs are presented in bottom panels (Panel \textbf{c} for ICME and Panel \textbf{d} for IP shock). The values of average and maximal errors are presented in the panels. Two additional \emph{dashed lines} illustrate deviation from the model of $\pm$10 hours.}
 \end{figure}

	It is also worth mentioning that in the case of fast CMEs (\emph{V}$>$1200~km~s$^{-2}$), the empirical model is a better
approach in predicting TTs. For these events, average errors are below five hours and the maximal error does not exceed ten hours.
This is illustrated in the figure by two additional dashed curves showing a deviation from the model of $\pm$ten hours.

 \subsection{CME Transit Time Estimation}

		Having determined the different initial velocities (maximal or average), we are able to calculate TTs directly using kinematic equations of motion. For this purpose, we assume that once the CME reaches the initial velocity it moves with an uniform negative acceleration. This movement is controlled by drag force due to interaction with the solar wind. Deceleration stops when the CME speed reaches that of the solar wind (i.e. 300\,\--400~km~s$^{-1}$). From this moment onwards the CME moves with constant velocity, which is recorded at the Earth [\emph{V}$_{\mathrm{FINAL}}$]. Knowing the acceleration and using the above assumptions, we can easily calculate TT. The data we have allows us to determine the effective acceleration of CMEs (\citealt{Gopalswamy01}). They are designated as the quotient of the speed difference determined in the vicinity of the Sun and that determined at the Earth's vicinity over the TTs of CMEs. It can be expressed by the equation
\begin{equation}
ACC_{\mathrm{EFFECTIVE}}=(V_{\mathrm{INITIAL}}-V_{\mathrm{FINAL}})/TT,
\end{equation}
 where $V_{\mathrm{INITIAL}}$ is velocity of CME determined near the Sun (average or maximal/\emph{V}$_5$) and  \emph{V}$_{\mathrm{FINAL}}$ is velocity of ICME measured at the Earth's vicinity and TT, is the transit time of a given CME that is defined in the previous sections.

	The relationship between initial velocities (\emph{V}$_{\mathrm{AVG-SOHO}}$, \emph{V}$_{5-\mathrm{SOHO}}$, \emph{V}$_{\mathrm{AVG-STEREO}}$,
\emph{V}$_{5-\mathrm{STEREO}}$) of CMEs and their effective accelerations are shown in Figure~7. The quadratic relationship between
the effective accelerations and the respective speeds are indicated by a solid line. These curves represent empirical models of
effective acceleration of CMEs depending on their initial speeds. \citet{Michalek04} demonstrated that the average
acceleration models cannot give accurate prediction of TTs. They proposed to introduce the effective acceleration of a CME which
is computed using a linear fit to two extreme samples of CMEs. The slowest events which have no acceleration and the fastest
events are accelerated up to 1 AU. These events are expected to have no acceleration cessation during their travel to
the Earth (\citealt{Michalek04}). The dashed lines, in Figure~7 are linear fits to the these groups of CMEs only.

 \begin{figure}
 \centerline{\includegraphics[width=1.0\textwidth,clip=]{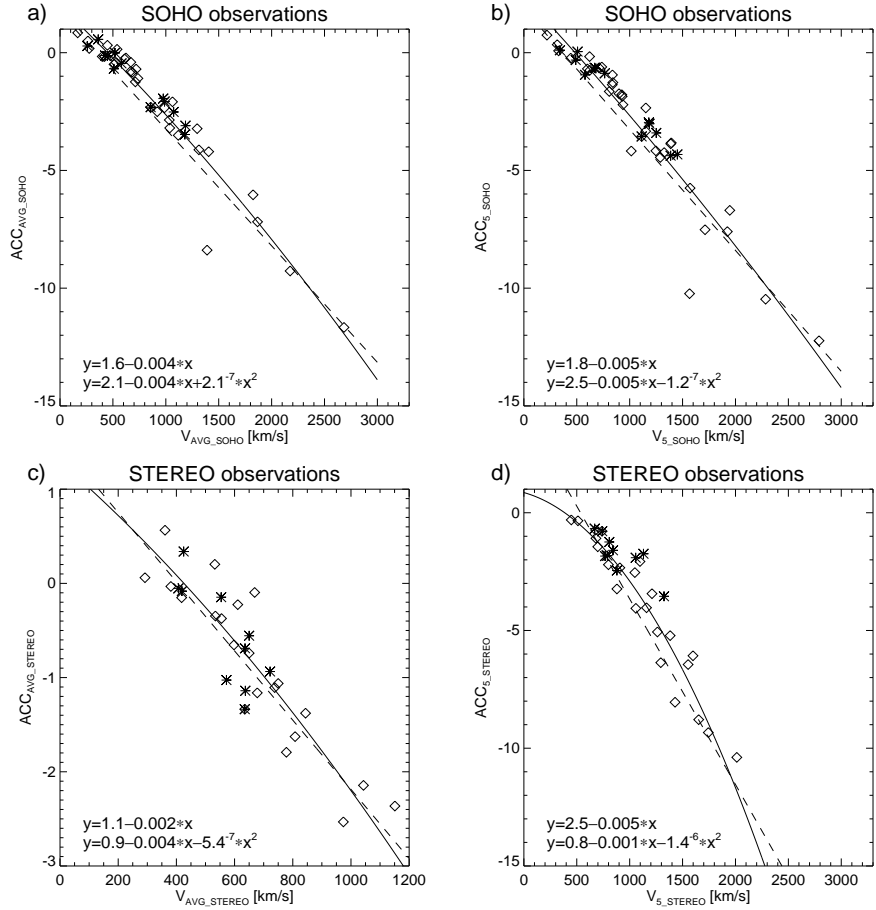}}
 \caption{Scatter plots of effective acceleration versus respective initial speed of CMEs [\emph{V}$_{AVG-SOHO}$, \emph{V}$_{5-\mathrm{SOHO}}$, \emph{V}$_{\mathrm{AVG-STEREO}}$, \emph{V}$_{5-\mathrm{STEREO}}$]. The \emph{open diamond symbols} are for non-interacting and \emph{star symbols} for interacting CMEs. The \emph{solid curves} are quadratic fits to the data points. The \emph{dashed lines} are linear fits for the three slowest and the three fastest events from the entire sample of CMEs.}
 \end{figure}

	With the help of different CME acceleration empirical models, we are able to determine the TT using general equations of motion.
The results are displayed in Figures~8 and~9. The scatter plots below display the observed $versus$ predicted TT for the respective
acceleration models displayed in Figure~7, which also shows a comparison of TTs for SOHO observations.
Unfortunately, the results are not very accurate. Average errors for models obtained from the average and maximal speeds are
above 13 and 15 hours, respectively.

\begin{figure}[h]
 \centerline{\includegraphics[width=1.0\textwidth,clip=]{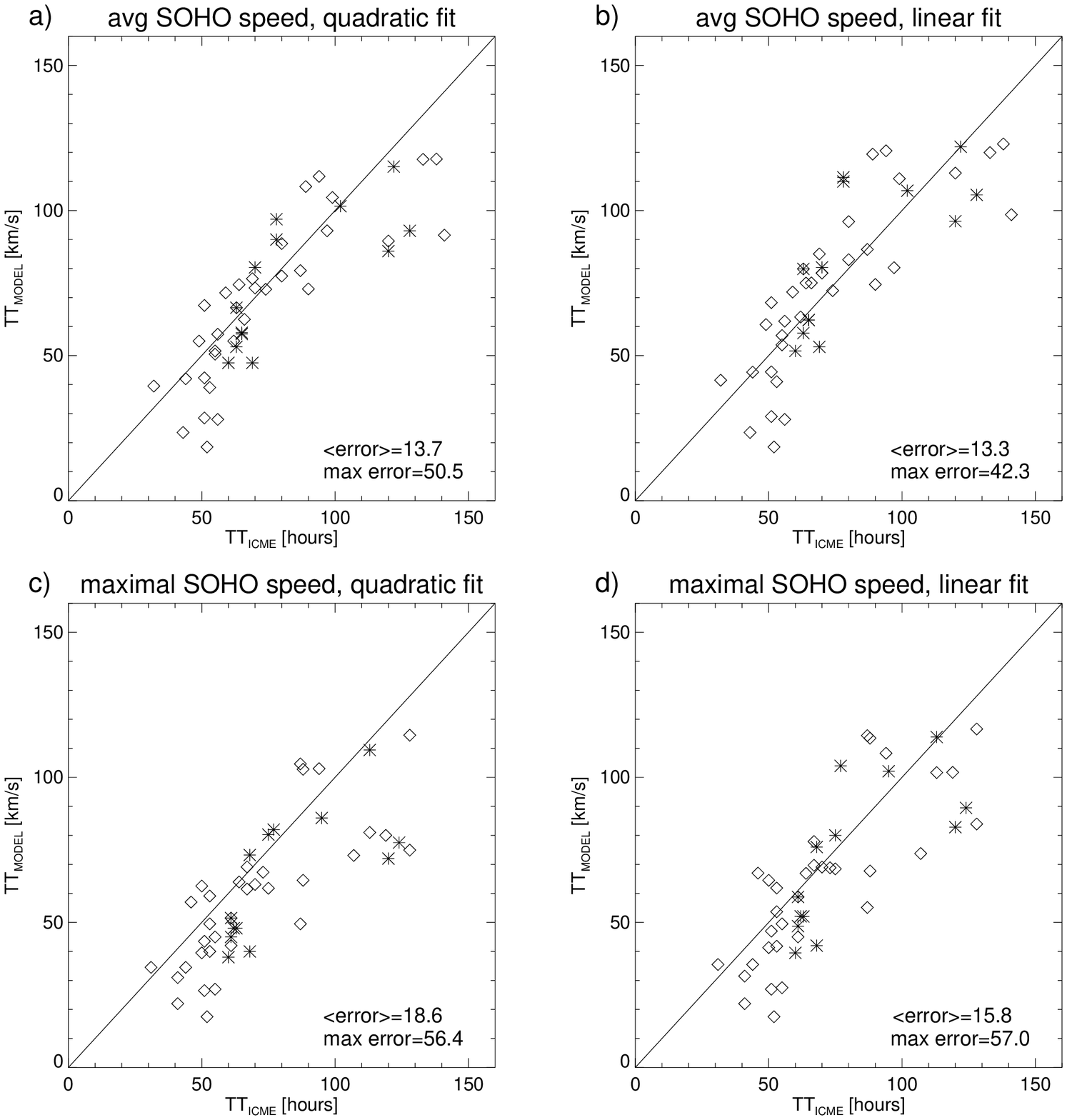}}
 \caption{Scatter plots presenting comparison between observed and predicted TT based on acceleration profiles obtained from SOHO observations. In the respective panels we have results for effective acceleration obtained from a quadratic fit (Panels \textbf{a} and \textbf{b}) and from a linear fit (Panels \textbf{c} and \textbf{d}). The \emph{upper panels} are for effective acceleration obtained from the average velocities and the bottom from the maximal/\emph{V}$_5$ velocities, respectively. The \emph{open diamond symbols} are for non-interacting and \emph{star symbols} for interacting CMEs. \emph{Solid lines} show the theoretical situation when observed and predicted times are identical.}
 \end{figure}

	As seen from the figure, the results seem to be better for effective accelerations determined from linear
fits (Panels b and d) in comparison to those received from quadratic fits (Panels a and c). For the effective
acceleration models obtained from the average velocities, data points are symmetrically scattered around the solid line,
that shows the ideal situation when both times are equal. In the case of the maximal velocities the data points are mostly
placed below this line. This means that on average the predicted TT are lower than observed TT.

 \begin{figure}[h]
 \centerline{\includegraphics[width=1.0\textwidth,clip=]{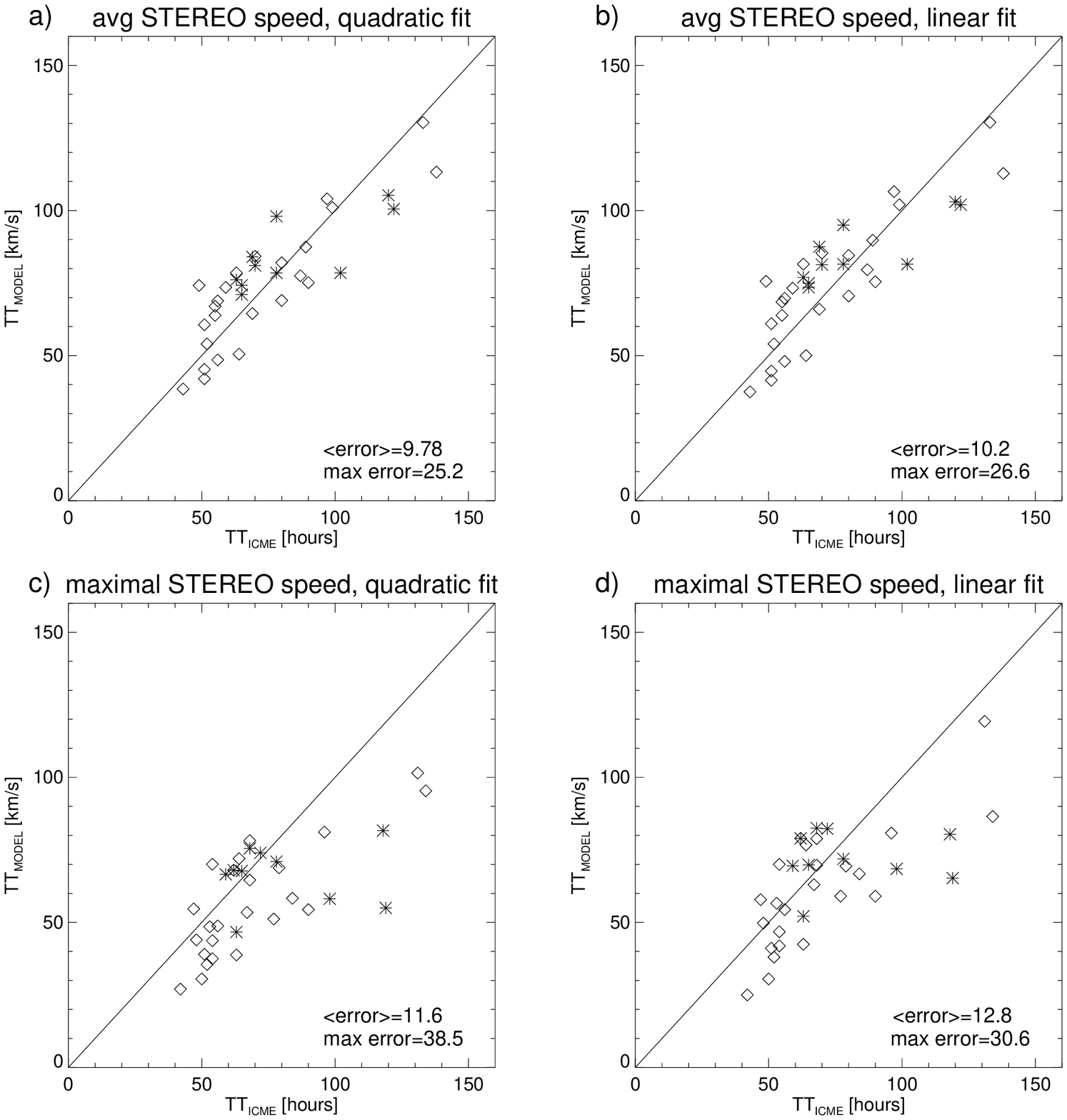}}
 \caption{Scatter plots showing comparison between observed and predicted TT based on acceleration profiles obtained from the STEREO observations. In the respective panels we have results for effective acceleration obtained from a quadratic fit (Panels \textbf{a} and \textbf{b}) and from a linear fit (Panels [\textbf{c} and \textbf{d}). The \emph{upper panels} are for effective acceleration obtained from the average velocities and the bottom from the maximal/\emph{V}$_5$ velocities, respectively. The \emph{open diamond symbols} are for non-interacting and \emph{star symbols} for interacting CMEs. \emph{Solid lines} show the theoretical model when observed and predicted TT are identical.}
 \end{figure}

	In Figure, we have similar comparisons but for STEREO observations. In this case the results are more promising.
For all of the models considered, the average errors are $\approx$9--12 hours. However, the best results are for the
effective acceleration obtained from the average velocities (Panels a and b). As seen from the figures, the average and
maximal errors are smallest and the symmetric scatter of data points around the solid line represents the ideal
situation when observed and predicted TT are equal. 

 \subsection{Errors $Versus$ the Range of CME Observation from the Sun}

	The results obtained show that the best accuracy in predicting TT are provided by models based on the average speed
determined from the STEREO observations (Figure 5a and Figure 9b). Using these two models the predicted TT are subject to
minimal (9.27 hours, 10.2 hours) and maximal (28.3 hours, 26.6 hours) errors, respectively.

 \begin{figure}[h]
 \centerline{\includegraphics[width=1.0\textwidth,clip=]{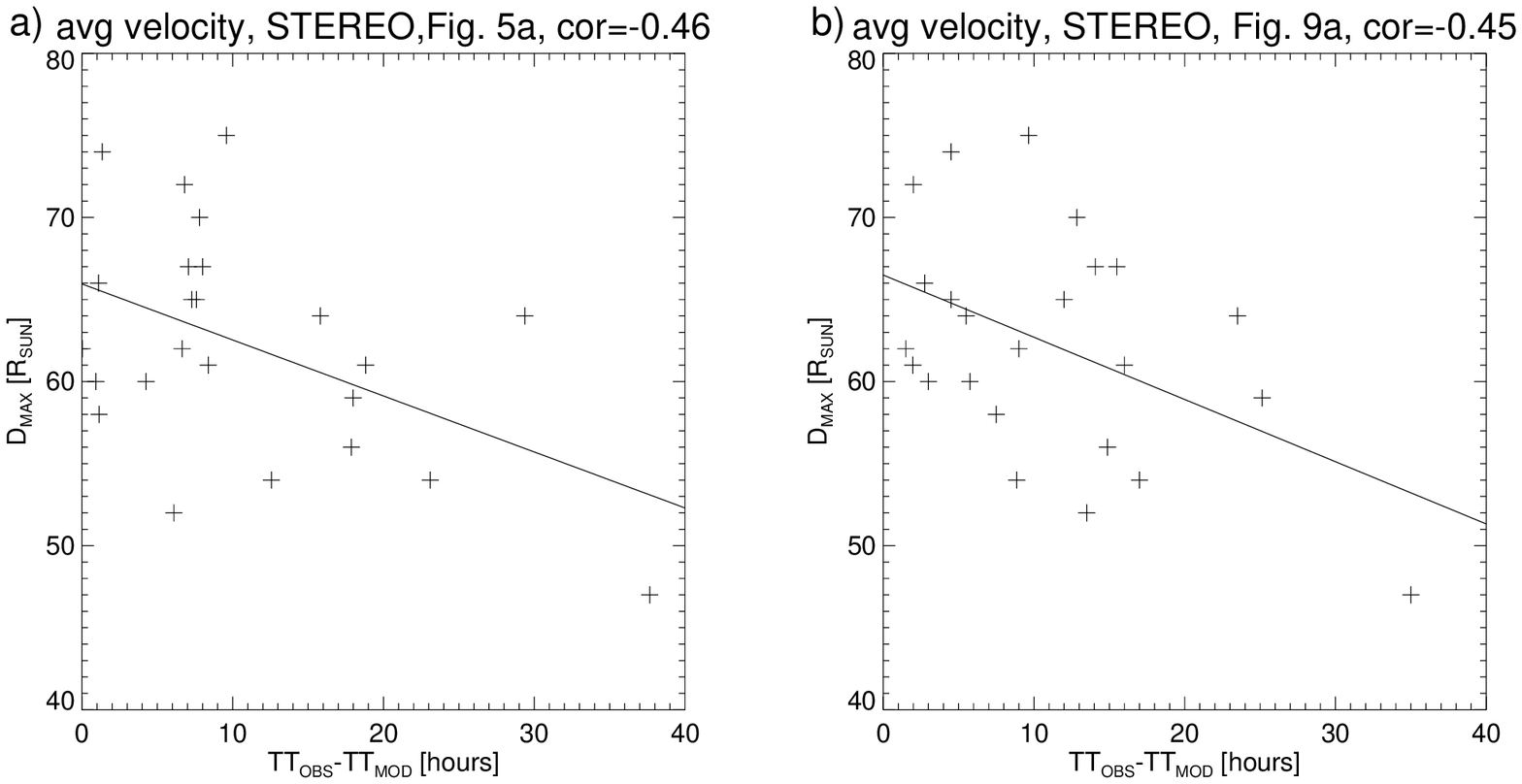}}
 \caption{Scatter plots of difference between predicted (errors of predictions) and observed TT versus range of observations of CME [D$_{\mathrm{MAX}}$]. The \emph{left panel} presents the results for the empirical model shown in Figure~5a and the \emph{right panel} presents the results for calculations displayed in Figure~9a.}
 \end{figure}

	It is important to consider why these observations provide the best results. Certainly, the direction of observation
has some influence. It seems that observations from the STEREO spacecraft, in the case of a CME directed toward the Earth,
are subject to  smaller projection effects than those conducted with the SOHO spacecraft. However, another important factor
to be considered is the field of view of the STEREO telescopes, which is much larger than the field of view of the  SOHO
coronagraphs. This effect is illustrated in Figure~10. This figure shows scattered plots of the difference between predicted
(errors of predictions) and observed TT $versus$ range of observations of CME [D$_{\mathrm{MAX}}$]. The parameter [D$_{\mathrm{MAX}}$] is the maximal distance where a given CME was recorded by the STEREO telescopes. The left panel presents the results for the empirical
model shown in Figure~5a and the right panel presents the results for calculation displayed in Figure~9a. As shown here,
the errors in prediction depends on the range of observations. The correlation coefficients are $\approx$0.5. This result is
consistent with considerations presented by \citet{Bronarska18}.
They showed that the errors in determining the speed depend, to the highest degree, on the number of height--time points.
The average speeds determined from the STEREO observation have the highest number of height--time points and are therefore
determined with the best accuracy.

 \section{Summary and Discussion}

	In this study we evaluate the TT of CMEs during the ascending phase of Solar Cycle 24. For this purpose we employed,
different from the previous studies, two additional STEREO coronagraphs observing the Sun in quadrature and provided a
new definition of initial speed of CMEs. It was demonstrated that all the considered initial speeds [\emph{V}$_{\mathrm{AVG-SOHO}}$,
\emph{V}$_{5-\mathrm{SOHO}}$, \emph{V}$_{\mathrm{AVG-STEREO}}$, \emph{V}$_{5-\mathrm{STEREO}}$] are significantly correlated, however, the most significant correlation appears to be between the velocities determined using the same spacecrafts. The maximal velocities
are larger by about 80$\,\%$ and 25$\,\%$ than the average velocities for the STEREO and SOHO telescopes, respectively.
The initial velocities determined in the STEREO images [\emph{V}$_{\mathrm{AVG-STEREO}}$, \emph{V}$_{5-\mathrm{STEREO}}$] are also significantly
related to the final velocities of an ICME. This is a very important result from the point of view of space weather.
This allows us to accurately predict one of the most important parameter determining geoeffectivness of CMEs.

	The TT of CMEs to the Earth is the next important parameter for space weather. 
We presented two methods, using the initial velocities of CMEs estimated near the Sun, to predict the TT. First, we used the correlation
between the initial velocities and the TT to generate empirical models predicting the TT. The best empirical models are for
the average speeds estimated in the STEREO images (Figure 4a). Second, we calculated the TT using kinematic equations of motions
and different models of effective acceleration. Again the best results were obtained for average velocities determined in the
STEREO field of view (Figure 9a).

	Using observations from the STEREO we were able to reduce the average absolute errors of TT prediction by only about an hour. However, what is more important is the fact that the new approach have radically reduced the maximum TT estimation errors
equal to 29 hours. Previous studies determined the TT with the maximum error equal 50 hours. We also tried to find out why the STEREO observations are more useful in determining the TT.
As the STEREO field of view is much larger than SOHO field of view, it  allows us to follow the CME up to one third of the way
to the Earth and thus more accurately determine their speed. It is shown that errors in predicting TT significantly depend
on the distance range of the CME observation, i.e. the larger the observation range, the smaller the error.

	In our research we defined a new initial velocity of a CME, the maximum velocity determined from the velocity
profiles obtained from a moving linear fit to five consecutive height--time points.
 This
new approach have radically reduced the
maximum TT errors equal to 29 hours. Previous studies determined the TT with the
maximum error equal 50 hours. Additionally,
the maximal velocities of CMEs are better correlated with the final ICME speeds in
comparison with previous models.
It is also worth noting that in the case of maximum speeds, the empirical model
obtained from the correlation between these  speeds and TT are in good agreement.  Results for these considerations are shown
in Figure 6. For the fast CMEs (\emph{V}$>$1200~km~s$^{-1}$), the empirical model works very well, as the average errors are below
five hours and the maximal error does not exceed ten hours.

	The model presented can be used universally. As input to this model, we can use speeds or accelerations obtained in
different ways. In particular, we can use the real three dimensional speeds estimated using stereoscopic observation from all
of the coronagraphs. It seems, however, that the results obtained in this way will not be significantly different from those
obtained from the STEREO observations. As we mentioned earlier, STEREO observations in quadrature, in the case of halo CMEs,
provide velocities very close to the real ones.

\vspace{1cm}
\textbf{Acknowledgements}
\\Anitha Ravishankar and Grzegorz Michalek were  supported by NCN through the grant UMO-2017/25/B/ST9/00536.\\

\section{Disclosure of Potential Conflicts of Interest}
The authors declare that they have no conflicts of interest.

%
%
%
%
%
%
%


\end{article}
\end{document}